\documentclass[]{IEEEtran}   

\usepackage{graphicx}
\usepackage[namelimits,intlimits,sumlimits]{amsmath}
\usepackage{amsfonts}
\usepackage{txfonts}
\usepackage{amssymb}
\usepackage{psfrag}

\begin{document}



\title{Analytical model of 1D Carbon-based Schottky-Barrier Transistors}
\author{Paolo Michetti, Giuseppe
  Iannaccone~\IEEEmembership{Member,~IEEE,}
  \thanks{This work was supported in part by the EC Nanosil FP7 Network of 
    Excellence under Contract 216171 and in part by the European Science 
    Foundation EUROCORES Programme Fundamentals of Nanoelectronics, through 
    fundings from Consiglio Nazionale delle Ricerche (awarded to 
    IEEIIT-PISA) and the European Commission Sixth Framework Programme under 
    Project Dewint (Contract ERAS-CT-2003-980409).}%
  \thanks{Authors are with the Dipartimento di Ingegneria dell'Informazione,
    Universit\'a di Pisa, Pisa I-56122, Italy (e-mail:
    p.michetti@iet.unipi.it; g.inannaccone@iet.unipi.it).}
}

\maketitle

\begin{abstract}
  Nanotransistors typically operate in far-from-equilibrium (FFE) 
  conditions, that cannot be described neither by drift-diffusion, nor by purely ballistic models.
  In carbon-based nanotransistors, source and drain
  contacts are often characterized by the formation of Schottky
  Barriers (SBs), with strong influence on transport.
  Here we present a model for one-dimensional field-effect transistors (FETs),
  taking into account on equal footing both SB contacts and FFE
  transport regime. 
  Intermediate transport is introduced within the B\"uttiker's probe approach to dissipative
  transport, in which a non-ballistic transistor is seen as a
  suitable series of individually ballistic channels.
  Our model permits the study of the interplay of SBs and
  ambipolar FFE transport, and in particular of the transition
  between SB-limited and dissipation-limited transport. 
\end{abstract}


\markboth{} {Michetti and Iannaccone: Analytical model of 1D SB transistors}

\section{INTRODUCTION}
Since the isolation of graphene in sheets~\cite{novoselov2004,novoselov2005},
with their exceptionally promising high mobility~\cite{geim2007},
graphene-related materials have attracted much
interest for their possible application in nanoelectronic devices.  
In particular, semiconducting carbon nanotubes (CNTs)~\cite{bachtold2001} and single-layer or
bilayer graphene nanoribbons (GNRs)~\cite{li2008} have been successfully employed in quasi-1D
nanotransistors.

An important issue related to carbon-based channels is the nature 
of the metallic contact at source and drain, which can lead to
different pinning of the Fermi level and
consequently to the formation of ohmic or Schottky contacts~\cite{javey2003,zhou2009}.
The presence of SB contacts can have dramatic effects on device
performance, because charge injection is
subordinated to a tunneling process.
However, in nanodevices with reduced oxide thickness, tunneling
phenomena at source and drain are favored, and, while they often limit
performance in conventional transistors, their
exploitation is at the core of the concept of tunneling FETs~\cite{knoch2008}.

Transport in nanotransistors is certainly far from of
equilibrium, but is still not fully ballistic, and currents are much lower than those
predicted by ballistic models~\cite{singh2006}.
While it is perfectly clear that inelastic scattering may arise from
the interaction of carriers with phonons and impurities, it is rather
complex to take into account microscopically its effect on transport.
A powerful phenomenological attempt to deal with carrier relaxation and
decoherence was based on the B\"uttiker virtual probes
approach~\cite{buttiker1985,buttiker1986}, in which inelastic
scattering is thought as localized in special points, spaced by a
length defined as ``mean free path''.
The Buttiker approach was also introduced in microscopical models based
on tight-binding Hamiltonians~\cite{damato1990}, and recently extended to
deal, via a quantum Langevin approach, with 1D conductors~\cite{roy2007}.     
In~\cite{wang2004} the B\"uttiker probes approach to inelastic scattering
was employed in a simulation, based on the non-equilibrium
Green's functions formalism, of a non-ballistic silicon nanowire
transistor.

Fully microscopical analysis of inelastic scattering due to specific mechanisms such as phonon scattering, with the non-equilibrium Green's 
functions approach, has also been addressed by adding
a proper self energy correction on a site-representation propagating Hamiltonian by Jin {\em et al.} \cite{Jin2006} and by M.~Gilbert
{\em et al.} \cite{gilbert2005,gilbert2007}.
 
As far as analytical models are concerned, transport in quasi-1D FETs
is generally treated as purely ballistic or with a drift-diffusion assumption as in Ref.~\cite{jimenez2003,jimenez2004,paul2007}.
A largely invoked approach to treat partially ballistic transport including the
effects of backscattering was proposed by Lundstrom et al.~\cite{lundstrom1997}. 
This approach, that is easily included as a correction to ballistic
models, has the merit of offering a very simple and synthetic
picture but does not allow a full description of the seamless transition from ballistic to quasi-equilibrium drift-diffusion transport.
Recently a rigorous semi-analytical model based on the B\"uttiker
virtual probes approach~\cite{buttiker1985,buttiker1986} has emerged, 
in which a non-ballistic transistor is seen as a
suitable chain of $N$ ballistic channels, where $N$ is the ratio of the channel length
to the mean free path, or equivalently as a series of
drift-diffusion and a ballistic FET~\cite{mugnaini2005,mugnaini2005b,michetti2009}.

In this work we propose a semi-analytical model based on the virtual
probes approach, which describes one-dimensional FETs, treating on
equal footing Schottky barrier contacts and FFE transport conditions.
In Section II we summarize the general analytical
description of graphene nanoribbons subbands, density of states,
equilibrium charge density, extensible also to the carbon
nanotube case. 
In Section III we present a WKB approximation of the tunneling
probability through Schottky barrier contacts, yielding analytical
expressions for the transmission based on two different levels of approximation for the
energy dispersion curves of GNRs (or CNTs). 
In Section IV a model for a single ballistic transistor
with SB contacts is presented, compared with data from numerical simulations. 
In Section V we propose a compact model, based on B\"uttiker
virtual probes approach able to deal with both intermediate transport
and SB contacts, and use it to study the interplay of SB and dissipative transport.    
\begin{figure}
  \centering
  \includegraphics[width=0.8\linewidth]{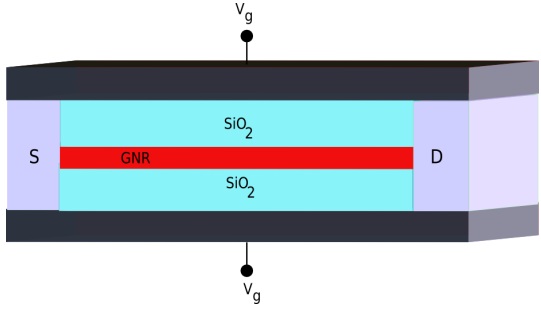}
  \caption{Sketch of a DG-GNR field effect transistor, considered as an exemple for the application of our model. 
  }
  \label{fig:gnr_sketch}
\end{figure}

\section{DISPERSION RELATION AND DENSITY OF STATES}\label{app2}
\par\noindent
The dispersion curve of an armchair GNR with $N$ dimer lines can be obtained analytically
by cutting techniques, analogous to that used for CNTs in~\cite{mintmire1998},
from the 2D graphene tight-binding dispersion.
The subband dispersion curves correspond to 1D segments of the graphene
Brillouin zone with the confined wavevector quantized as
$k_\alpha=\frac{\pi\alpha}{N+1}$, with $\alpha=1,2,\dots,N$.
The dispersion curve of the subband $\alpha$, referred to midgap, is
\begin{equation}
  E_\alpha(k) = \pm V \left\{ 1 + 4
  \cos{\frac{\sqrt{3}ak}{2}}A_\alpha +
  4A_\alpha^2 \right\}^{1/2},    
  \label{eq:energyGNR}
\end{equation}
with $A_\alpha=\cos{\left(\frac{\pi \alpha}{N+1}\right)}$.
We note here that a dispersion relation totally analogous to (\ref{eq:energyGNR}) applies to zig-zag
($N$,$0$) CNT, with the only difference that in the place of
$A_{\alpha}$ we have to use $A_\alpha^{CNT}=\cos{\left(\frac{\pi
    \alpha}{N}\right)}$, where $\alpha$ is the subband
index of CNTs~\cite{akinwande2008}.

\begin{figure}
  \centering
  \includegraphics[width=0.8\linewidth]{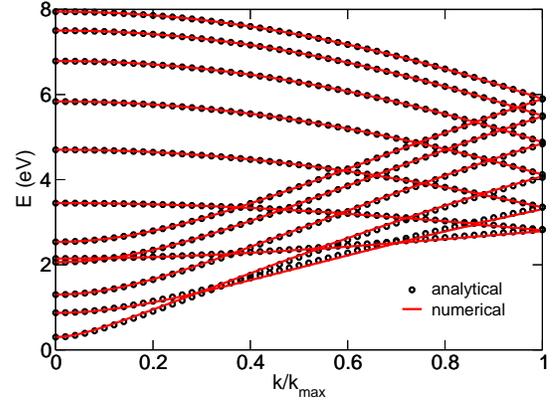}
  \caption{Comparison of the subbands of a A-GNR with $12$ dimer
    lines between a numerical tight-binding calculation and
    our analytical result with edge corrections. Valence bands are symmetrical. 
  }
  \label{fig:numericoemio}
\end{figure}

Therefore, much of the results for GNRs obtained here and in the following of
the paper, with the exclusion of the edge corrections, can be directly
generalized to the zig-zag CNT case.
The edge of the $\alpha$-th subband $E_\alpha(0)$ is expressed as
\begin{equation}
  E_\alpha(0) = \pm V \left( 1+2A_\alpha \right). 
  \label{eq:Ecut}
\end{equation} 
Let us note that the lowest lying subband is given by the value of
$\alpha$ for which $A_\alpha +\frac{1}{2}$ is minimum. 
The edges of the nanoribbon are laterally exposed to vacuum
and experience a different chemical environment, 
therefore the hopping parameter between carbon atoms at the edges tends to be slightly different.
We can, at least partially, account for the presence of edges via a
perturbative approach to the first order~\cite{son2006}.
The perturbation theory to the first order leads to the following
eigenenergy corrections: 
\begin{equation}
  \delta E_\alpha(k) = (\pm)_\alpha H^{ed}_{\alpha,\alpha} = (\pm)_\alpha \frac{4 v}{N+1} \sin^2{\left(\frac{\alpha
      \pi}{N+1}\right)} \cos{\left(k a_{c-c}\right)},
\label{eq:correction} 
\end{equation}
with $v=0.12$~eV the energy correction of the hopping parameter at the edges in the tight
binding Hamiltonian.
The correction has a positive or negative contribution depending on the wavefunction
parity with respect to the two asymmetric carbon atoms, which are
connected by the edges.
Therefore if  $A_\alpha\ge-\frac{1}{2}$ we have a
positive contribution $(\pm)_\alpha=1$, otherwise a negative one $(\pm)_\alpha=-1$.
The edge-corrected energy dispersion relation, which we will refer to as
the full band (FB) approximation when applied to FET modelling, is therefore
\begin{equation}
  E_\alpha^c(k) = E_\alpha(k) +\delta E_\alpha(k).    
  \label{eq:energy_edge}
\end{equation}
The comparison between numerical tight-binding calculations, with edge effects
taken in account, and the analytical result with perturbative corrections, for a A-GNR
of $12$ dimer lines is shown in Fig.\ref{fig:numericoemio}.
The agreement is very good, especially at $k=0$, where (\ref{eq:energy_edge}) 
reproduces the results of~\cite{son2006}.
For simplicity we define here the band edges as $\varepsilon_\alpha=E_\alpha^c(0)$. 

\subsection{Approximated expressions}
\par\noindent 
In modelling nanotransistors only the
lowest laying subbands matter, in which the relevant transport phenomena take place. 
For these lowest lying subbands often an effective mass (EM)
approximation is invoked  
\begin{equation}
  E_\alpha^{EM}(k) = \varepsilon_\alpha + \frac{ \hbar^2 k^2}{ 2 M_\alpha},    
  \label{eq:dispEM}
\end{equation}
in this case the following effective mass for the $\alpha$-th mode can been employed
\begin{equation}
  M_\alpha = -\frac{2}{3}\frac{\hbar^2\varepsilon_\alpha}{a^2V^2A_\alpha}.
  \label{eq:massEM}
\end{equation}
The DOS in EM approximation is given by 
\begin{equation}
  D_\alpha^{EM}(E) = \frac{2}{\pi \hbar} \sqrt{\frac{M_\alpha}{2E}}, 
  \label{eq:densitaEM}
\end{equation}
with $E$ expressing the `kinetic energy', i.e. the energy calculated
with respect to the band edge $\varepsilon_\alpha$.

The EM approximation is rather crude, and an intermediate (I)
approximation, between the FB and the EM, can be the use
of the dispersion curve
\begin{equation}
  E_\alpha^I(k) = \pm \sqrt{ \varepsilon_\alpha^2 + \frac{\varepsilon_\alpha\hbar^2k^2}{M_\alpha}},    
  \label{eq:dispI}
\end{equation}
for which the DOS is
\begin{equation}
  D_\alpha^I(E) = \frac{2(\varepsilon_\alpha +E)}{\pi \hbar}
  \sqrt{\frac{M_\alpha}{\varepsilon_\alpha E
  \left(E + 2\varepsilon_\alpha \right)}}. 
  \label{eq:densitaI}
\end{equation}
\begin{figure}
  \centering
  \includegraphics[width=0.9\linewidth]{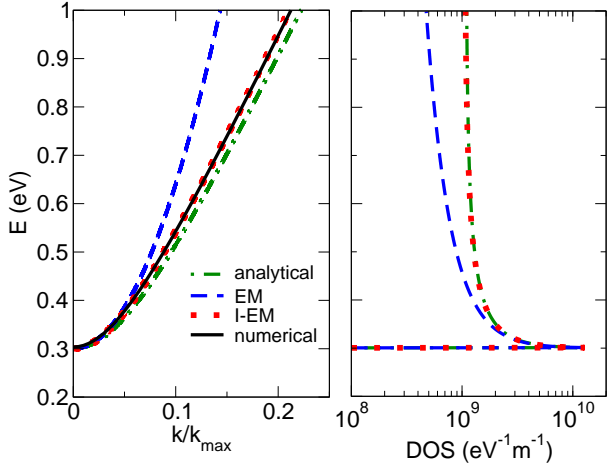}
  \caption{Energy dispersion curve, and the corresponding density of states, of the lowest conduction subband of a
    A-GNR with $12$ dimer lines.
    A numerical tight binding result is compared with our FB analytical result and
    with EM and I approximate dispersions.
    In the energy range considered here the agreement between
    numerical, FB and I approximations is excellent. 
  }
  \label{fig:DISPR}
\end{figure}
In Fig.\ref{fig:DISPR} we compare the lowest band dispersion curve and
the corresponding DOS for a GNR with 12 dimer lines.
Both the FB and the I approximations reproduce quite well numerical
tight-binding calculations, and give similar DOS, of course the I
dispersion is only accurate for energies $E\ll V$.
The EM approximation instead remains quite accurate only for about $E<0.1$~eV.

\subsection{Carrier density}
\par\noindent
The carrier density affects both electrostatics and transport properties.
Here we develop a similar analysis to what done in~\cite{akinwande2008} for
carbon nanotubes.
The electron carrier density per subband can be expressed as
\begin{equation}
  n_\alpha=\int_{0}^{\varepsilon_\alpha^{top}-\varepsilon_\alpha}
  f(\frac{E+\varepsilon_\alpha-q\phi_c-\mu}{kT})D_\alpha(E) dE,
  \label{eq:carrier_density}
\end{equation}
where $\mu$ is the Fermi level, $f(x)=(1+\exp{x})^{-1}$ the
Fermi-Dirac distribution, and $\varepsilon_\alpha^{top}$ is top edge
of the $\alpha$-th subband, that for most purposes
can be taken as $\infty$, due to the finite extension of $f(E)$.
$\phi_c$ is the electrostatic potential in the device, which rigidly
shift the levels.
Because the non-negligible contribution to (\ref{eq:carrier_density}) comes from states near
$\varepsilon_\alpha$, we can use the intermediate expression for 
the DOS $D^I_\alpha(E)$. 
If we consider a non-degenerate situation ($\varepsilon_\alpha-3kT>\mu$),
typical of sub-threshold regimes in FETs, we obtain
\begin{equation}
  n_\alpha= \frac{2\sqrt{M_\alpha (\varepsilon_\alpha-q\phi_c)}}{\pi
  \hbar} e^{\beta (q\phi_c+\mu)}
  \int_1^\infty e^{-\beta (\varepsilon_\alpha-q\phi_c) z} \frac{z}{\sqrt{z^2-1}} dz,
  \label{eq:carrier_density2}
\end{equation}
with $z=E/\varepsilon_\alpha$ and $\beta=(kT)^{-1}$.
With a partial integration and recognizing the modified Bessel
function of the second kind $K_1$, the charge density can be
expressed as
\begin{equation}
  n_\alpha= \frac{2}{\pi
  \hbar} \sqrt{M_\alpha \varepsilon_\alpha} e^{\beta (q\phi_c+\mu)} K_1(\beta \varepsilon_\alpha).
  \label{eq:carrier_density3}
\end{equation}
In order to give an estimation of the Bessel function $K_1$ which has
no closed form, we can adopt the approximation~\cite{akinwande2008} 
\begin{equation}
  K_1(x)\approx \frac{K_{1/2}(x)+K_{3/2}(x)}{2}=\sqrt{\frac{\pi}{2x^3}}\frac{1+2x}{2}e^{-x}
\end{equation}
arriving in the end to express the charge density as
\begin{eqnarray}
  n= N_c e^{-\beta(\varepsilon_\alpha-q\phi_c-E_F)}, \\
  N_c=\sqrt{\frac{M_\alpha}{2\pi\beta^3}} \frac{1+2\beta
  \varepsilon_\alpha}{\hbar \varepsilon_\alpha},
\end{eqnarray}
with essentially the same form of 3D bulk semiconductors.

\section{TUNNELING OF SCHOTTKY BARRIERS}
\par\noindent
Our aim is to provide an analytical description of the tunneling through SB contacts.
The first step is to model in the simplest way the
potential decay occurring near the source and drain contacts.
The potential inside a transistor channel is described by the a 3D Poisson equation 
\begin{equation}
  \nabla^2 \phi(\vec{r})=- \frac{\rho(\vec{r})}{\epsilon}
\end{equation}
together with the boundary conditions enforced by voltages $V_s$, $V_d$, $V_g$ at the source, drain
and gate leads.
In the evanescent mode analysis approach the electrostatic potential
inside a nanotransistor $\phi(\vec{r})$ is thought as the sum of a long-channel
solution $\phi_L(\vec{r})$, which satisfies
the vertical electrostatics, 
plus a short-channel solution $\phi^*(\vec{r})$, called evanescent mode, responsible of the 
potential variation along the channel~\cite{oh2000}. 
The short-channel solution is obtained, solving the Laplace equation
for the device with an adequate
expansion in harmonic functions.
As a matter of fact the short-channel solution
near the source contact results in an exponential profile
\begin{equation}
  \phi^*(\vec{r})\propto R(\vec{r}_\|) e^{-z/\lambda}.
\label{eq:scphi}
\end{equation} 
where $R(\vec{r}_\|)$ describes the solution in the channel cross
section and $\lambda$ comes to be a the natural scale length for the potential
variation in the device. 
The actual value of $\lambda$ depends on the details of the device
geometry, however in double-gate (DG) configuration, and considering that in general,
in carbon-based FET, the oxide
thickness is significantly larger than the channel thickness, the
asymptotic value $\lambda=(2t_{ox}+t_{ch})/\pi$ can be assumed.
In the case of a cylindrical GAA-CNT FET,  an explicit
calculation of $\lambda$ via evanescent mode analysis has been performed in~\cite{hazeghi2007}.

We follow this line and assume  that the channel potential rigidly shifts the confinement
eigenvalues $\varepsilon_\alpha$, where $\alpha$ runs on the different
subbands.
Now we are interested only in the potential inside the restricted zone
of the graphene channel $\phi_c(z)$, in which it can be assumed as a
constant (which is strictly true in subtreshold regimes), and we
consider its variation only along
the channel direction.
The long channel solution inside the channel is reduced to $\phi_L(\vec{r})\approx\phi_c$, 
where $\phi_c$ is solely imposed by
the vertical electrostatics, while the short-channel solution has the
form (\ref{eq:scphi}).
Therefore the potential in the channel $\phi_c(z)$
can be expressed as 
\begin{equation}
  \phi_c(z) = \phi_c + \frac{A_s}{q} e^{-z/\lambda}
    \label{eq:barrier_gen}
\end{equation}  
with $\phi_c=\phi(\infty)$ fixed by the vertical electrostatics and
$A_s$ imposed by the boundary condition at the SB contact
$A_s=E_{SB}^{(s)} -\varepsilon_L +q\phi_c$, where $L$ refers to the
lowest lying subband, due to the
Fermi level pinning at the metal/semiconductor interface.
$E_{SB}^{(s)}$ is the Schottky barrier height on the first conduction subband
with respect to the source Fermi level.
The charge injected from the source with energy lower than the barrier have
to tunnel in order to reach the channel.
We need to calculate the transmission through an
exponential decaying barrier of the kind
\begin{equation}
  E_{SB}(z) = A_s e^{-z/\lambda}
  \label{eq:barrier_simple}
\end{equation}   
with the height $A_s$ dependent on the electrostatic potential $\phi_c$. 
We note that however that if the band bending exceeds the energy gap
$2\varepsilon_\alpha$, carrier with energy
$0<E<A_s-2\varepsilon_\alpha$, 
will experience a SB of an height $A_s=E+2\varepsilon_\alpha$.

In order to estimate the behavior of a nanotransistor it is essential to
accurately describe tunneling phenomena, both in traditional FETs and in TFETs.
In this section we compare the tunneling calculated with WKB approximation in a full band approach
(FB-WKB), within the effective mass approach (EM-WKB) and 
intermediate approximation (I-WKB).
FB-WKB is more complex to implement and requires a numerical
solution of the integral
\begin{equation}
  ln(T(E))=-2\int_{z_1}^{z_2}\mathcal{I} [k_z(z;E)]dz
  \label{eq:int_WKB}
\end{equation}
While for the others two an analytical expression for the tunneling
$T(E)$ can be obtained.

\subsection{Effective-mass WKB approximation}
The transmission coefficient obtained via WKB approximation is given as
\begin{equation}
  T(E) = \begin{array}{ll}e^{-2\int_{z_1}^{z_2}
    \sqrt{2m_\alpha/\hbar^2 \left(E_{SB}(z)-E \right)}dz};
    \hspace{0.2cm} E<A_s
    \\ 1; \hspace{0,2cm} E\ge A_s
\end{array}
\end{equation} 
where $z_1,z_2$ are the classical turning points are   
\begin{equation}
  z_1 = 0  \nonumber; \hspace{0.5cm} z_2 = -\lambda \ln \left[ \frac{E}{A_s}\right]. 
  \label{eq:turning}
\end{equation}
The transmission coefficient can be analytically calculated in
\begin{equation}
  \ln{T(E)}=\hspace{-0.03cm} -\hspace{-0.03cm}4\hspace{-0.03cm}\lambda \hspace{-0.03cm}\sqrt{\frac{2\hspace{-0.03cm}m_\alpha
  \hspace{-0.03cm}(A_s-E)}{\hbar^2}} \hspace{-0.03cm}\left[\hspace{-0.1cm} 1 \hspace{-0.03cm} -\sqrt{\frac{E}{A_s\hspace{-0.03cm}-\hspace{-0.03cm}E}}\tan^{-1}\hspace{-0.03cm}\left(\hspace{-0.03cm}\sqrt{\frac{A_s\hspace{-0.03cm}-\hspace{-0.03cm}E}{E}}\right)\hspace{-0.03cm} \right].
  \label{eq:tunnel}
\end{equation}

\subsection{I WKB approximation}
Let us consider a dispersion curve of the kind (\ref{eq:dispI}).
The turning points with a barrier like (\ref{eq:barrier_simple}) are
the same as (\ref{eq:turning}), but now, under the barrier, the
imaginary part of the wavevector as a function
of the energy is given by 
\begin{equation}
  \mathcal{I}[k_z,E]=\sqrt{\frac{M_\alpha}{\hbar^2  \varepsilon_\alpha}} \sqrt{a^2-(b-e^{-\frac{z}{\lambda}})^2}
\end{equation}
with 
\begin{equation*}
  a=\frac{\varepsilon_\alpha}{A_s};\hspace{0.5cm}   b=a+\frac{E}{A_s}.
\end{equation*}
The integration (\ref{eq:int_WKB}), for $E<A_s$, leads to the WKB tunneling probability 
\begin{eqnarray} 
  \ln{T(E)} = \frac{2A_s\lambda\sqrt{M_\alpha}}{\sqrt{\hbar^2\varepsilon_\alpha}} \left[
    -b\left(\frac{\pi}{2}-\arctan{\frac{b-1}{R_1}}\right)-R_1 +\nonumber\right.\\
    \left.+R_2\left(\pi-\arctan{\frac{R_1 R_2}{a^2-b^2+b}} \right)\right]
  \label{eq:T_I}
\end{eqnarray}
where we introduced the abbreviations
\begin{equation*}
  R_1 = \sqrt{a^2-(b-1)^2};\hspace{0.5cm} R_2 = \sqrt{b^2-a^2}.
\end{equation*}

\subsection{Full-band WKB approximation}
For an armchair GNR, subband dispersion curves
are in the form (\ref{eq:energyGNR}), from which we can express the
wavevector as a function of the energy as 
\begin{equation}
  k=\frac{2}{a\sqrt{3}}\arccos{x},
  \label{eq:k}
\end{equation}
with the substitution $u=-z/\lambda$ and normalizing all quantities to
$A$, $x$ given by 
\begin{equation}
  x = \frac{ ( \frac{E}{A_s} + \delta - e^u)^2 - \alpha^2}{\nu},
  \label{eq:z}
\end{equation}
where we introduced
\begin{equation}
  \delta = \frac{\Delta}{A_s} \hspace{0.5cm}
  \alpha = \frac{(1+4A_\alpha)V^2}{A_s^2} \hspace{0.5cm}  \nu =
  \frac{4A_\alpha V^2}{A_s^2} \nonumber.
\end{equation}
In the integration domain of (\ref{eq:int_WKB}), the argument $x$ of the
inverse cosine function has module larger than $1$, and therefore
\begin{equation}
  \mathcal{I}(k)= -\frac{2}{a\sqrt{3}}\ln{\left|z+\sqrt{z^2+1}\right|},
  \label{eq:k_im}
\end{equation}
leading to the WKB tunneling probability 
\begin{equation}
  \ln{T(E)}=-\frac{4\lambda}{a\sqrt{3}} \int_{0}^{\ln{E/A_s}} \ln{\left|x+\sqrt{x^2+1}\right|} du.
  \label{eq:int_WKB_fin}
\end{equation}
\begin{figure}
  \centering
  \includegraphics[width=0.9\linewidth]{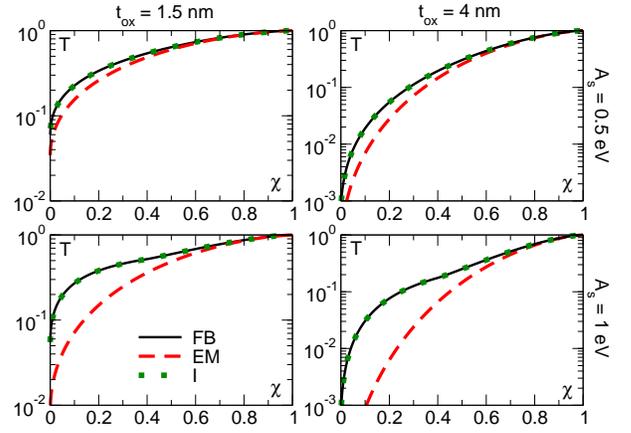}
  \caption{Transmission probability for an electron in the lowest
    laying subbands of a A-GNR with $12$ dimer lines.
    The FB result comes from the numerical integration of (\ref{eq:int_WKB_fin}),
    while we have analytical transmission probability in I (\ref{eq:T_I}) and
    EM (\ref{eq:tunnel}) approximation. 
  }
  \label{fig:tunn}
\end{figure}

In Fig.\ref{fig:tunn} we compare the tunneling coefficients, calculated
with the EM, I and FB WKB approaches, for a SBs of height $0.5$ and
$1$~eV, and for a $\lambda$ typical of DG A-GNR with $t_{ox}=1.5$ and $4$~nm.
Essentially the intermediate approximation completely reproduces the FB
tunneling probability, while a significant deviation is observed
with the EM-WKB approximation for $E< 0.5 A_s$.
Therefore the intermediate approximation seems an optimal
approximation for compact models in order to reduce the computational
times retaining high accuracy.

\section{SCHOTTKY BARRIER BALLISTIC FET}
\par
\begin{figure}
  \centering
  \includegraphics[width=0.9\linewidth]{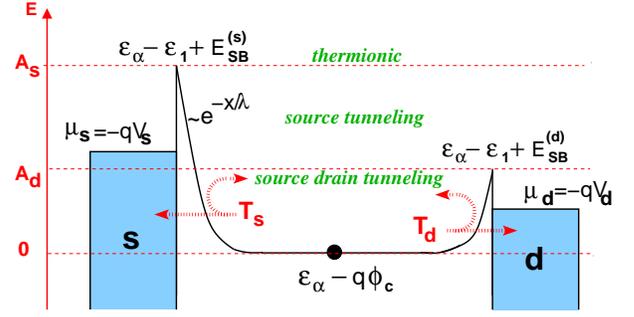}
  \caption{Conduction band edge profile of a SB
    nanoscale FET. The thermionic and tunneling energy ranges are shown.
  }\label{fig:barriers}
\end{figure}

\noindent
We consider here a ballistic transistor with Schottky barrier contacts
at source and drain, as shown in Fig.\ref{fig:barriers}.
As usual in compact models, we assume a complete phase randomization along the channel,
neglecting phase resonances in the transmission probability of the two
tunneling barriers, while multiple reflection events are taken into account.
Between two tunneling barriers, the forward and backward distribution
functions are modified by the multiple elastic
scattering~\cite{buttiker1988,hazeghi2007}.
The overall mobile charge, given by the sum of forward and
backward going charge carriers in the channel, can be expressed as
\begin{equation}
  \frac{Q_i}{q}\hspace{-0.03cm} = \sum_\alpha \hspace{-0.05cm}\int_{0}^{\varepsilon_\alpha^{top}-\varepsilon_\alpha}\hspace{-0.05cm} dE D_\alpha(E)
     \bigg\{  \frac{T_s\hspace{-0.03cm} \left(2\hspace{-0.03cm} -\hspace{-0.05cm} T_d \right)}{T^*}f(\eta_{\alpha,s}^i)  
    +  \frac{T_d\hspace{-0.03cm} \left(2\hspace{-0.03cm} -\hspace{-0.05cm} T_s \right)}{T^*}f(\eta_{\alpha,d}^i) \bigg\}
  \label{eq:chargeSD}
\end{equation} 
with $i=e,h$ for the electron and hole charge, where 
\begin{eqnarray}
  \eta_{\alpha,s(d)}^e = \frac{E -q\phi_c +\varepsilon_\alpha-\mu_{s(d)}}{k_b T}
  \label{eq eta_e}\\
  \eta_{\alpha,s(d)}^h = \frac{\mu_{s(d)}-E +q\phi_c -\varepsilon_\alpha}{k_b T},
  \label{eq eta_h}
\end{eqnarray}
and  
\begin{equation}
  T^* = T_s+T_d-T_sT_d,
\end{equation}
where $T_s$, $T_d$ are the tunneling coefficients at source and
drain, depending on both energy and channel potential. 
In order to compute the channel potential $\phi_c$, and, through it, the
subband energies, the total mobile charge $Q=Q_h-Q_e$ must be equal to
the charge induced by the electrostatic
coupling of channel with gate, source and drain through the
capacitances $C_g,C_s,C_d$ respectively:
\begin{equation}
  Q(\phi_c) = -\sum_{i=g,s,d,}C_i( V_i -V_{FB,i} -\phi_c),
  \label{eq:charge}
\end{equation} 
where $V_{FB,i}=\phi_i-\chi_g$ is the flat band voltage, given by the
difference beween the contact workfunction and the graphene electron
affinity.

The current is obtained with the Landauer-B\"uttiker formalism, which,
accounting for the tunneling, takes the following form
\begin{equation}
  I_i(\phi_c) = \frac{q}{\pi\hbar} \sum_\alpha \int_0^{E_\alpha^{top}-\varepsilon_\alpha} \frac{T_s T_d}{T^*}[f(\eta_{\alpha,s}^i)-f(\eta_{\alpha,d}^i)]dE.
  \label{eq:ISD}
\end{equation}
with $i=e,h$ accounting for the current of electrons and holes, and the
total current given by $I=I_e-I_h$. 
We note that (\ref{eq:chargeSD}) and (\ref{eq:ISD}) include both tunneling and
thermionic contributions.

\begin{figure}
  \centering
  \includegraphics[width=0.9\linewidth]{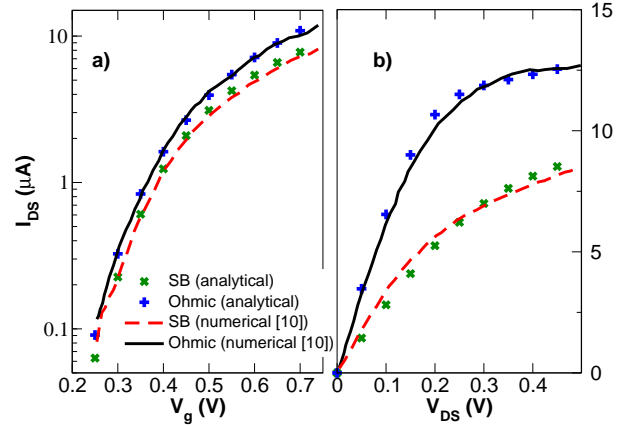}
  \caption{Comparison between our model and numerical simulations
    from ref.\cite{yoon2008}. 
    The transfer characteristics (a) at $V_{ds}=0.5$~ V and
    characteristics (b) at $V_g=0.75$~V of a  ballistic double-gate armchair GNR FET, with Ohmic and
    SB contacts of height $E_g/2\approx0.3$~eV are shown. 
    Assuming a GNR thickness of about $1$ nm we obtain $\lambda \approx 1.3$ nm. }
  \label{fig:fiori13}
\end{figure}

We apply our model to the case of a double gate armchair graphene
nanoribbon transistor (DG A-GNR FET) with both Ohmic and Schottky barrier contacts.
In Fig.\ref{fig:fiori13} we compare the transfer-characteristics (a)
and the output characteristics (b) of a ballistic armchair GNR FET, obtained
with our model and with numerical simulations based on the non-equilibrium Green's
function formalism in Ref.~\cite{yoon2008}.
The SiO$_2$ gate oxide thickness is $1.5$~nm, the armchair GNR
lattice is characterized by $12$ dimer lines, which correspond to a width of
$1.35$~nm and a bandgap of $0.6$~eV.
We employed here the intermediate analytical description of the GNR
subbands and density of states (\ref{eq:densitaI}).
The source and drain capacitances $C_s,C_d$ are introduced because of the short-channel
nature of the GNR simulated in~\cite{yoon2008} and are fixed, with
respect to the gate capacitance $C_g=1.1\times10^{-10}$~F/m, to $C_s=C_d= 0.1C_g$.    
%
The agreement between the numerical simulations and our compact model, for both curves
(Fig.\ref{fig:fiori13}) with ohmic and SB contacts, is very good,
demonstrating that the effects of SBs are well accounted for.

\section{SB TRANSISTORS IN INTERMEDIATE TRANSPORT REGIME}
To describe dissipative transport, we follow the approach developed in~\cite{mugnaini2005,mugnaini2005b} for a 2D
MOSFET for the non-degenerate and degenerate cases, and in~\cite{michetti2009} for quasi-1D FETs.
Such treatment is here expanded to include ambipolar devices. 
We recall that, within the B\"uttiker probes approach, inelastic
scattering is thought as localized in special points, spaced by a
length defined as ``mean-free path'' $\ell$.
The virtual probes act as localized reservoirs along the channel, in which
carriers are fully thermalized in equilibrium with the probe quasi-Fermi
energy $\mu_n$.
Transport from one virtual probe to the next is considered purely ballistic.
We have a drift-diffusion transistor when the channel length is much
longer than the free mean path, that from our point of view it is
equivalent to have a long enough chain of ballistic transistors, as rigorously shown in~\cite{mugnaini2005}.
On the contrary, when the number of internal contacts is small, transport is far-from-equilibrium, and fully ballistic in the limit $N=1$.
\begin{figure}
  \centering
  \includegraphics[width=0.8\linewidth]{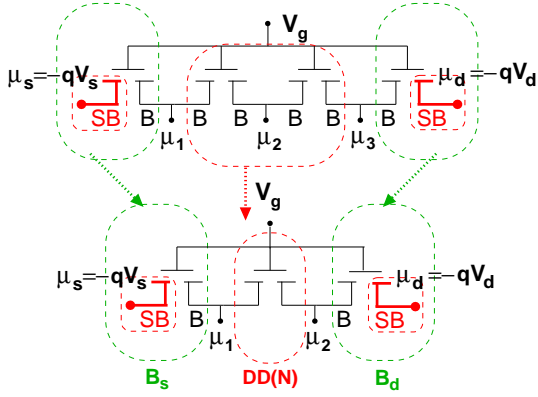}
  \caption{Chain of $N$ ballistic transistors with SB
    contacts at source and drain (first and last contact).  
    As explained in the text, the chain of ballistic transistors can
    be described as the series made by a central DD section accounting for dissipative
    transport in the $N-2$ internal nodes and by head and tail ballistic
    transistors accounting for the SB contacts with source and drain. 
  }
  \label{fig:modello}
\end{figure}

A transistor with SB contacts in the FFE transport regime is therefore
modeled as a series of individually ballistic channels, connected by fully thermalizing virtual probes placed at $x_{n}=n\ell$ with
$n=1,\dots,N-1$, with electrochemical potential $\mu_n$.
Head and tail of the series are connected to source
and drain through SB contacts as sketched in Fig.\ref{fig:modello},
and boundaries are fixed as $\mu_0=\mu_s$ and $\mu_N=\mu_d$. 
In the $n$-th ballistic channel $\mu_{n-1}$ and $\mu_{n}$ act as source and
drain, simultaneously solving (\ref{eq:chargeSD}) and
(\ref{eq:charge}) we can fix the channel potential
$\phi_c^{(n)}$.
In the same manner the current in the $n$-th channel is obtained with
(\ref{eq:ISD}), imposing  $\mu_{n-1}$ and $\mu_{n}$ as source and
drain Fermi levels. 
Since the current $I_n$ in any $n=1,\dots,N$ FET must be equal
to $I_{ds}$, we have $N$ equations determining the local Fermi
energies $\mu_n$.
We note that a distinction between ballistic internal channels ($B$) and boundary
channels with source ($B_s$) and drain ($B_d$) can be made.
In fact the first and the last ballistic channels are
characterized by SB contacts with metallic source and drain, while 
internal channels, in the region between the fictitious virtual
probes, can be treated as ohmic transistors.
The numerical solution of the complete chain of $N$ elements: 2
of the boundary kind and $N-2$ of internal kind, will be addressed as the $B(N)$ model.

Now we note that for the internal part of the chain the analysis developed in
\cite{michetti2009} applies.
In particular it has been shown that the current in an ohmic-contact
ballistic chain of $N$ elements, after a linearization procedure, can be
arranged in a to a drift-diffusion-like form (that we refer as the
$DD(N)$ model) in which the current is calculated through the formula 
\begin{equation}
  I_{ds} = \frac{q^2 \Gamma(1)\ell}{\pi\hbar L} \sum_\alpha \int_{V_s}^{V_d}
  \{ F_{-1}(\eta^e_\alpha[V])-F_{-1}(\eta^h_\alpha[V])   \} dV,
  \label{eq:Ids_chain}
\end{equation}
where $F_{-1}(x)$ is the Fermi-Dirac integral of order $-1$, $\Gamma$ the gamma function and 
\begin{eqnarray}
  \eta^e_\alpha=(q\phi_c-qV-\varepsilon_\alpha)/kT,\\
  \eta^h_\alpha=(-q\phi_c+qV+\varepsilon_\alpha)/kT.
\end{eqnarray}
We note that $\eta$, not only directly depends on $V$, but also
indirectly through $\phi_c$, which is self-consistently imposed by
the linearized vertical electrostatics
\begin{equation}
  \begin{array}{ll}
    Q_m = C_g(V_g-V_{FB}-\phi_c[V])\nonumber\\
    Q_m = -\frac{q\Gamma(1/2)}{\pi}\sum_\alpha 
    \sqrt{\frac{2k_bTm_\alpha}{\hbar^2}} \{ F_{1/2}(\eta^e_\alpha[V])-F_{1/2}(\eta^h_\alpha[V]) \}.
  \end{array}
\end{equation}
The linearized $DD$ model (\ref{eq:Ids_chain}) has also the advantage
of dealing  with non integer $N=L/\ell$, and is therefore more flexible than
the ballistic chain itself. 
As noted in~\cite{michetti2009}, (\ref{eq:Ids_chain}) can be
rearranged in a local form, analogous to a DD equation $I_\alpha=\mu_\alpha Q_\alpha
\frac{dV}{dx}$,  where the degenerate mobility (we consider now a
monopolar regime) is given by 
\begin{equation}
  \mu_\alpha^e= \frac{qv_\alpha\ell}{2kT} \frac{F_{-1}[\eta_\alpha^e]}{F_{-1/2}[\eta_\alpha^e]}
\end{equation}
with $v_\alpha=\sqrt{\frac{2kT}{\pi m_\alpha}}$ the mean carrier velocity.
This expression gives us a link between $N=L/\ell$ and the mobility.

We can now model a SB transistor in intermediate transport
regime as a series of $B_s$-$DD(N)$-$B_s$ segments, with two nodes
between the boundary channels and the internal segment, characterized
by electrochemical potentials that can be fixed
exploiting the current continuity in the device.
We will refer to this macro-model as the $BDDB(N)$ model.
This compact model permits to analyze both the presence of Schottky
barrier contacts and far-from equilibrium transport condition, while
keeping low the computational burden, especially with respect
to numerical simulations including dissipation.

\begin{figure}
  \centering
  \includegraphics[width=0.9\linewidth]{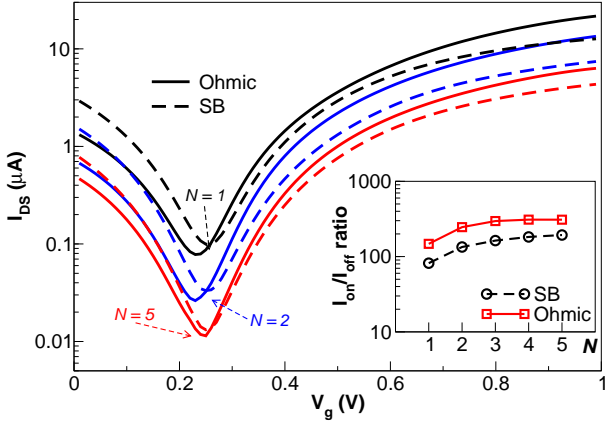}
  \caption{Transfer characteristics of a ballistic chain made of a
    series of $N$ DG GNR FETs, with $N=1,2,5$ and $V_{ds}=0.5$~V,
    calculated with our model. 
    In the inset the $I_{\rm on}/I_{\rm off}$ ratio
    for $V_g^{(\rm off)}=0.25$~V and $V_g^{(\rm on)}=0.75$~V as a function of $N$.
    %
  }
  \label{fig:fiori1b}
\end{figure}

We now analyze the effects of inelastic scattering on
the performance of a DG A-GNR FET.
In non-ballistic transport (increasing $N$) the transfer characteristics
(Fig.\ref{fig:fiori1b}) vertically shift, in a semilog plot, as
expected due to the mobility reduction.
It is interesting to note that the effect is more marked in the subthreshold region and,
consequently, an increase of the $I_{\rm on}/I_{\rm off}$ ratio as a function of $N$ is
observed, as shown in the inset.
%
In ballistic models with positive $V_{ds}$, in subthreshold
regime, tunneling from the drain leads to hole
accumulation under the channel, which increases the quantum
capacitance and reduces the control over channel.
Subsequently a larger subthreshold swing and a lower $I_{on}/I_{off}$ is
obtained.

\begin{figure}
  \centering
  \includegraphics[width=0.9\linewidth]{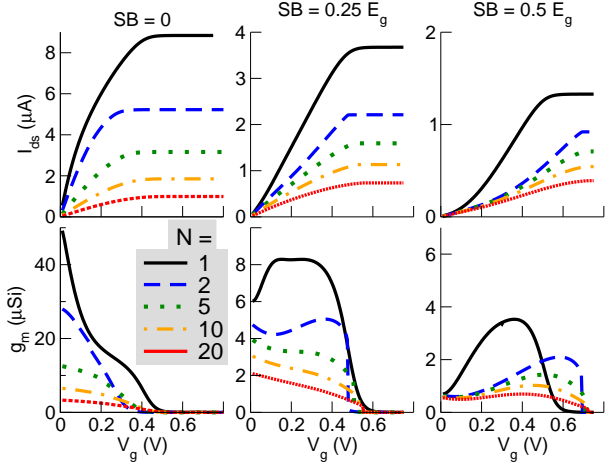}
  \caption{Output characteristics and differential conductance for of
    a GNR ballistic chain made of $N$ DG GNR FETs, with $N=1$, $2$,
    $5$, $10$, $20$.
    Three different pinning of the SB with respect to the conduction
    band are considered: without Schottky barrier (SB=0), with a SB of $E_g/4$ and
    $E_g/2$.
    The details of the device are the same as Fig.\ref{fig:fiori13},
    except $t_{ox}=5$~nm, for which all the features due to the SB are
    enhanced due to the thicker barrier.
  }
  \label{fig:g_m}
\end{figure}

%
An accurate analysis of the SB effects on output characteristics can be
performed calculating the differential conductance
$g=\partial I_{ds}/\partial V_{ds}$.
In fig~\ref{fig:g_m} we compare the output characteristics and the
differential conductance for a device with $t_{ox}=5$~nm,
with a SB height SB$=0$, $0.25$, $0.5$ $E_g$. 
Note that the presence of Schottky barrier contacts is more relevant in transistors with a
looser vertical confinement, where the tunneling barriers are thicker.
We observe that in samples with SB$=0$~eV the output characteristics concavity is always negative,
and the differential conductance is monotonously decreasing with
$V_{ds}$.
If the SB height is finite the differential conductance acquires a
non-monotonous behavior, which well describes the ``S'' shaped concavity change of the
characteristics curves before reaching saturation, especially evident
in thicker SB devices.  
It is interesting to note as, apart from a reduction of the maximum saturation
current, larger ballistic chains (larger $N$), in which a higher
inelastic scattering is active, lead to a smoothening of the
non-monotonous dependence of $g$ on $V_{ds}$.
In this fact we can recognize a gradual transition between devices in
which the characteristics are dominated by SB contacts and 
devices in which inelastic relaxation is predominant.    

\begin{figure}
  \centering
  \includegraphics[width=1\linewidth]{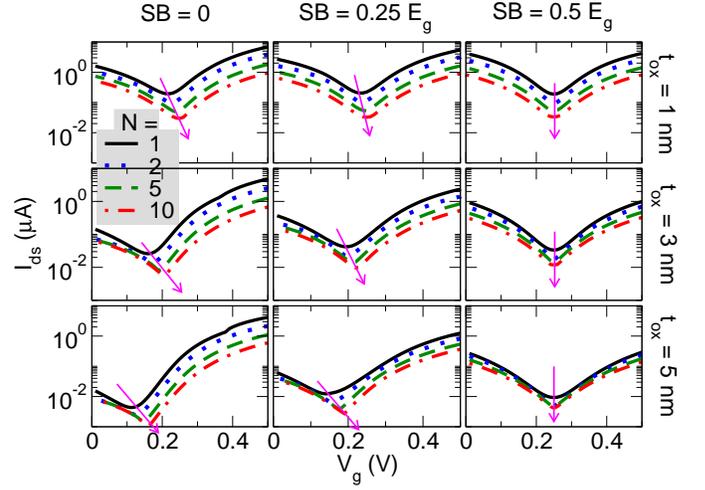}
  \caption{Transfer characteristics of GNR devices increasing
    $t_{ox}=1$, $3$, $5$~nm, calculated for $V_{ds}=0.5$~V.
    Ballistic chains of $N=1$, $2$, $5$, $20$ are drawn, for devices without 
    Schottky barrier (SB=0), with a SB of $E_g/4$ and
    $E_g/2$ are shown.
    Arrows indicating the shift of the transfer characteristics
    curves with $N$ are also added as a guide for the eyes.
  }
  \label{fig:tox}
\end{figure}

In electron-hole symmetrical materials as undoped graphene nanoribbons
or carbon nanotubes, the relative SB height with respect to the
bandgap determines the position of the minimum of transfer
characteristics, it influences their shape and their symmetry (see Fig.\ref{fig:tox}).   
A SB of height $E_g/2$ preserves the bandstructure electron-hole
symmetry and therefore results in transfer characteristics
which span symmetrically from the current minimum off state (placed at $V_g=V_{ds}/2$).
Curves calculated reducing the SB height for electrons (for $E_g/4$
and $0$) show a growing asymmetry,  with weaker hole currents and larger
electron currents, together with a shift of the transfer
characteristic minimum to lower values of $V_g$.   
This phenomenon is prominent in thicker SB devices such as the
$t_{ox}=5$~nm FET, but well observable
also in a $t_{ox}=1$~nm device. 
The increase of the lateral confinement leads in fact to an almost linear
increase of the SB thickness and therefore all tunneling processes
become harder.
As expected, if we increase the dissipative phenomena (increasing $N$)
a reduction of the current is observed.
But more interesting, while the SB$=E_g/2$ curves vertically shift
along the segmented line, the shift of the
other curves is diagonal, note in fact the horizontal shift
of their minima with $N$.
Moreover, increasing $N$, the minima seem to converge towards
the value $V_g=V_{ds}/2$, typical of a symmetrical ambipolar  device.
This is yet another signature of the growing importance of inelastic
transport over the SB contacts.
Therefore, for sufficiently well-confined FET, we can expect in quasi-ballistic GNT/CNT devices to
clearly observe a SB behavior, which become more and more subtle in dissipative regimes.           
To quantify the relative importance of the Schottky barrier in
determining the symmetry of the transfer characteristics we
made the following physical estimation:
SB$=0.5E_g$ corresponds to the symmetrical case, therefore if we impose a
different SB the change in the conductance will be exponential in the SB difference 
$\delta E_{\rm{SB}}$ as 
\begin{equation}
  \delta g_{\rm SB} \propto \exp{\left\{-\frac{2t_{ox}}{\pi \hbar} (2 m
  \delta E_{\rm{SB}})^{1/2}\right\}}
\end{equation}
as can be obtained estimating the differential
conductance of a device with a SB source contact at the source Fermi
level.
This quantity is in fact dominated by the tunneling coefficient (\ref{eq:tunnel}).
This difference in the conductance is relevant as long as it is
greater than the conductance due to the DD(N) chain.
We obtain
\begin{equation}
  \gamma = \frac{\delta g_{\rm{SB}}}{g_N} \approx N \exp{\left\{-t_{ox}*\frac{2}{\pi
  \hbar} (2 m \delta E_{\rm{SB}})^{1/2} \right\}}
\end{equation} 
Employing this formula we can calculate the $N=N_s$ corresponding to
$\gamma=1$ for different SB value and oxide thickness, as shown in the
following table
\begin{center}
  \begin{tabular}{l|lll}
    \centering
    $N_s$ &$t_{ox}=1$ & $t_{ox}=3$ &  $t_{ox}=5$~(nm)  \\
    \hline
    SB$=0$       &    $4$         &   $70$      &    $10^3$  \\
    SB$=0.25E_g$ &    $7$         &   $450$     &    $10^4$
  \end{tabular}
\end{center}
$N_s$ gives a rough estimation to the number of nodes (i.e. $L/\ell$
ratio) needed to make the
transfer characteristics symmetrical, in spite of the presence of a SB. 
As can be observed comparing these values with the behavior of curves
in Fig.\ref{fig:tox}, the $t_{ox}=1$~nm curves with SB$=5$ and $10$,
respectively for SB$=0.25E_g$ and $0$, are quite
symmetrical in accordance with $N_s=4$ and $N_s=7$ found by our
calculation.
The minimum of the curve $N=10$ with SB$=0.25E_g$ comes near to the
symmetrical values, but still misses it being our estimation $N_s=70$. 
Other curves are highly asymmetric being $N\ll N_s$.

\begin{figure}
  \centering
  \includegraphics[width=0.9\linewidth]{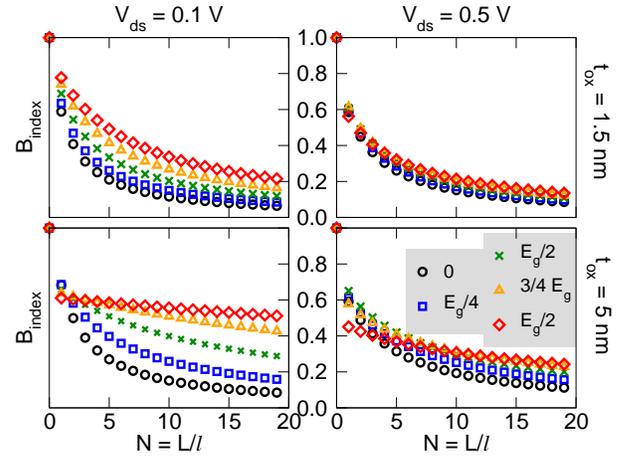}
  \caption{Ballistic index of a ballistic chain of $N$ elementary GNR
  FETs with SB is varied from $0$ to $E_g/2$ as a function of $N$.
  The source-drain voltage is set to $V_{ds}=0.1$~V and $0.5$~V
  for $V_g=0.75$~V. 
  }
  \label{fig:Bindex}
\end{figure}

A typical parameter used to characterize the transport regime in
quasi-ballistic devices is the ballisticity index $B_{index}=I/I_1$,
which is the ratio of the actual current to the current
corresponding to an analogous device in a purely ballistic transport
regime ($N=1$).
In Fig.\ref{fig:Bindex} we analyze the role of the SB contacts in
determining the ballisticity index as a function of $N$, and therefore
as a function of the degree of
inelastic relaxation. 
In general to lower SB heights correspond a faster variation of the
$B_{index}$ with $N$, with a sudden drop of the ballisticity as function of the
number of nodes, after which a slower decrease is observed.
SBs affect in particular the ballisticity index calculated for lower
$V_{ds}$, due to the concavity of the output characteristics,  
while larger source-drain voltages reduce the relative importance of
SB with respect to inelastic mechanisms.
Calculations with $t_{ox}=5$~nm reveal the increased importance of SB
contacts, and reflect the presence of the inflection in the output
characteristics, with a concavity change before saturation.
In particular, for higher value of the SB we observe a slower
dependence of the $B_{index}$ on $N$, because the
current is calculated in a bias point of the characteristic curve of strong
``s'' curvature.
Physically, it means that the current flowing in the device is mostly limited by the injection
through the tunneling barriers.    

\section{CONCLUSIONS}
We have presented a semi-analytical model dealing with ambipolar
one-dimensional
Schottky barrier transistors in intermediate transport regimes between
fully ballistic
and quasi equilibrium, i.e. governed by the drift-diffusion model.
We have introduced simplified, but accurate, descriptions of the Schottky
barrier profiles and of the electrostatics, and
analytical approximations of the tunneling coefficients of
the Schottky barriers.
We demonstrate that a Schottky barrier transistor can be modeled as
three transistor is series,
with common gate voltage. The central one is a drift-diffusion transistor,
with mobility dependent on the degree of degeneracy of the
one-dimensional carrier gas.
The other two transistors are ballistic FETs with a Schottky barrier
contact corresponding to the
external actual contacts (source or drain).
In the case of ballistic transport, our model allows us  to reproduce
the results of a
3D numerical Poisson-Schroedinger simulator. In the case of very long
channel, with respect
to the mean free path, current is limited by the central
drift-diffusion transistor.
The model allows very directly to investigate the transition from
barrier-limited transport to  channel-limited transport.
Our semi-analytical model represents an accurate and
simple way to gain physical insights into the behavior
of nanoscale transistors with Schottky barrier contacts, including
most the relevant physics at a very low computational cost.

\bibliographystyle{IEEEtran}
\bibliography{IEEEabrv,bibf}

\end{document}